# Magnification by Gravity

—A gravitationally lensed type Ia supernova may yield a precise and accurate measurement of the Hubble constant.

*Chien-Hsiu Lee (李見修), Subaru Telescope, National Astronomical Observatory of Japan*
*650 North A'ohoku Place, Hilo, HI 96720, USA; email: leech@naoj.org*

In the era of precision cosmology, the idea is to determine the expansion rate, or Hubble constant ($H_0$), to the one per cent level using the cosmic distance ladder — that is, the chain of methods based on Cepheids and type Ia supernovae (SNe), for example, to build up estimates of increasing astronomical distances — and compare it with cosmic microwave background (CMB) measurements. This approach can provide constraints on the equation of state of dark energy, the masses of neutrinos and the spatial curvature of the Universe. Using the distance ladder, the SH0ES team[1] has measured $H_0$ to 2.4%, in tension (at the 3.4σ level) with the latest CMB results from the Planck satellite. Although it is tempting to claim this discrepancy as being due to systematics in the Planck measurement or a sign of new physics, it is important to have independent, local $H_0$ measurements with uncertainties comparable to the cosmic distance ladder. One alternative and single-step approach to measure $H_0$ is by measuring the time delay introduced by gravitational lensing with multiply imaged quasars[2]. Recently, Ariel Goobar and colleagues[3] discovered a multiply imaged, gravitationally lensed type Ia SN — iPTF16geu (Fig. 1), a promising stepping-stone to probe the mass distribution in the lensing galaxy, breaking lensing degeneracies and further pinning down $H_0$.

According to Einstein's theory of general relativity, a foreground galaxy can induce spacetime curvature and bend the lights from a background source to form multiple images, where each of the images will go through different light paths (geodesics). As the difference of the light travel time (or time delay) only depends on the spacetime curvature — which is a combination of $H_0$ and the gravitational potential from the intervening mass — we can therefore measure $H_0$ as long as we have an accurate mass model in hand. Supernovae were first proposed to measure the time delay[4]; however, due to the lack of all-sky transient surveys in previous decades, time-delay measurements have been carried out using quasars instead, which suffer from systematic errors due to lensing degeneracies in the lens modelling. This uncertainty stems from our not knowing the intrinsic luminosity of the quasars, and the only observables from multiply imaged quasars are the image positions, the magnification ratio between images and the time delays. However, all these observables remain unchanged if we add a sheet of constant mass density to the mass model[5]. And worse still, this additional sheet of mass can either be internal to the lensing galaxy, or originate from masses along the line of sight. Thus, in order to better constrain the mass model, we will need to conduct extensive spectroscopic follow-ups of the lensing galaxy and its environment[6]. The picture changes completely if the background source is a standard candle, from which the image magnification factor can be derived. Then, the magnification factor[7], in contrast to the magnification ratio, is strongly dependent on the mass profile, and hence can be used to directly break the mass-sheet degeneracy. As the intrinsic luminosity of type Ia SNe is well-known, we can thus derive the magnification factor. Moreover, with the distinct light-curve shape of type Ia SNe, we can also obtain the time-delay with high accuracy.

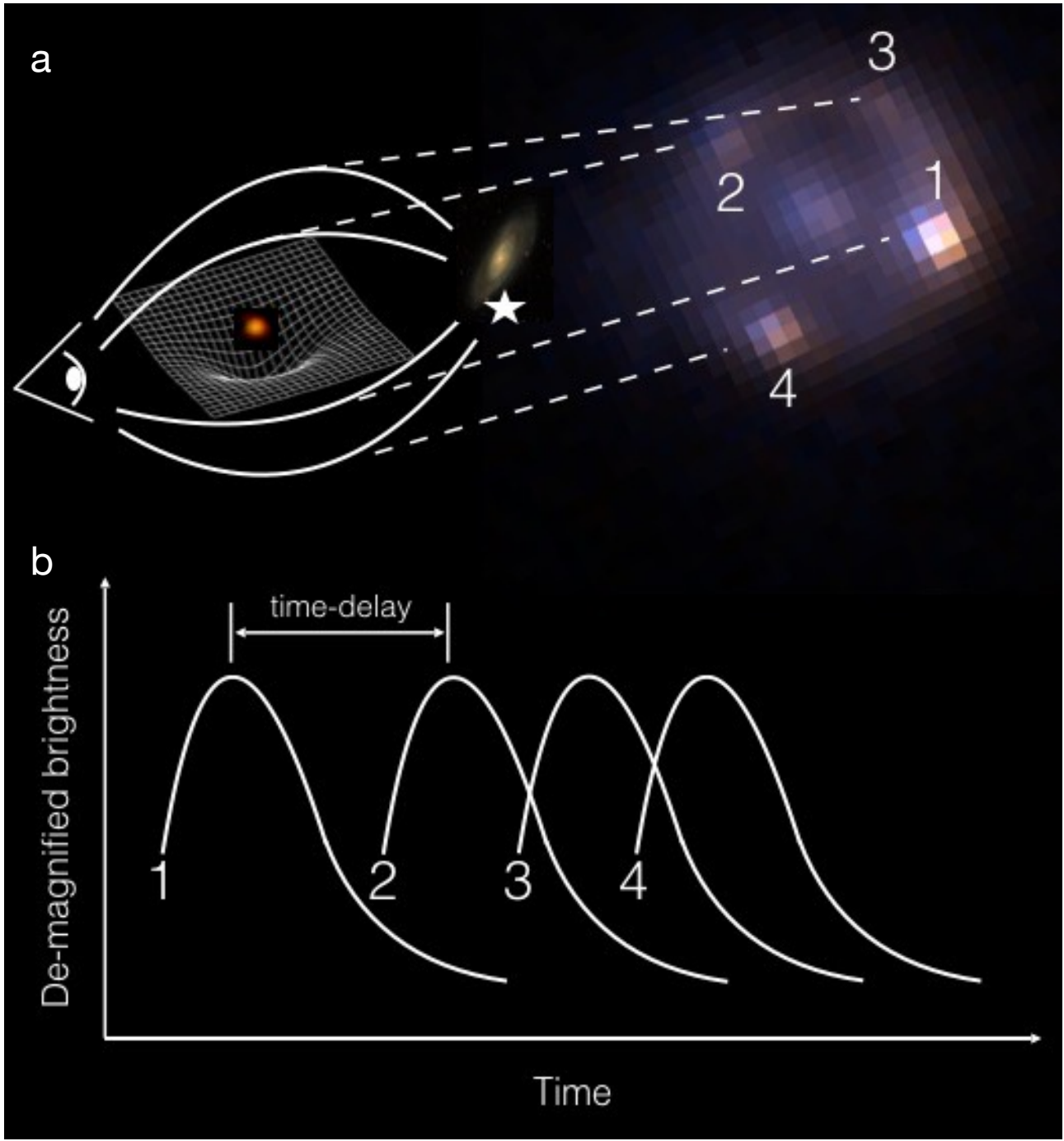

Fig. 1. Gravitational lensing of type Ia supernova iPTF16geu. **a**, The spacetime between iPTF16geu (marked with a star symbol) and the observer (on Earth) is disturbed by the gravity of the lensing galaxy (in orange). The observer will see the host galaxy of the supernova form a ring-like structure in the background, and the supernova split into four images. **b**, As type Ia supernovae have a distinct light-curve shape, we can easily determine the time delay between the four images (shown in the plot of demagnified brightness as a function of time, with the same numbering as in **a**).

With the advent of wide-field cameras and large-area surveys, it is now possible to identify gravitationally lensed SNe. For example, the Pan-STARRS survey has discovered PS1-10afx, a highly magnified (by a factor of 30) type Ia SN, which could have been multiply imaged[8]. However, as the lensing nature was established long after the supernova faded, there lacked timely high-resolution imaging to confirm the multiply imaged nature during the explosion. The first multi-image supernova, SN Refsdal[9], was found with exquisite spatial resolution delivered by the Hubble Space Telescope (HST) during a high cadence survey of massive clusters, but it is a type II SN, which prevents us from accurately measuring $H_0$. Making use of the high cadence, large area, automated and real-time transient classification and follow-up enabled by the intermediate Palomar Transient Factory (iPTF) survey[10], Goobar *et al.* were able to identify iPTF16geu as a candidate lensed type Ia SN before it faded away, and promptly triggered high-resolution imaging follow-up from ground-based adaptive optics and the HST, resolving the four lensed images and establishing the multiply imaged nature of the SN. Furthermore, Goobar and co-workers also triggered high cadence (~3 days), optical and near-infrared photometric monitoring with the HST. This includes rest-frame I-band monitoring, which will enable us to catch the second bump during the fading phase of the light curve and determine the time delay.

However, gravitationally lensed type Ia SNe still suffer from other caveats, such as substructures in the lens galaxy. Indeed, when calculating the event rate from the iPTF survey, Goobar *et al.* conclude that high magnification events such as iPTF16geu are very rare, unless we underestimate the substructure in the lens galaxy. In addition, the magnification differences among the four images cannot be well explained by the smooth-lens model, suggesting that iPTF16geu is subject to substructures in the lens galaxy. Whereas Goobar *et al.* estimate that the timescale of such an effect from an isolated, point-mass object is relatively long (>9 years) and should not affect the shape of the light curve, previous studies by Dobler *et al.*[11] showed that a network of point-mass objects might alter the shape of the light curve and prevent accurate time-delay measurements. With the high cadence HST follow-up, we will soon find out whether gravitationally lensed type Ia SNe such as iPTF16geu are capable of providing a precise and accurate estimate of $H_0$.